\documentclass[preprint,usenatbib]{mn2e}
\usepackage[normalem]{ulem}
\usepackage{graphicx}
\usepackage{appendix}
\usepackage{amsfonts} 
\usepackage{ulem} 
\usepackage{color} 
\usepackage{parskip} 
\usepackage{hyperref}

\newcommand{\D}{\Delta}
\newcommand{\e}{\epsilon}
\newcommand{\dt}{\Delta_{threshold}}
\newcommand{\s}{\sigma}
\renewcommand{\S}{\Sigma}

\newcommand{\nf}{f_{false}^{\rm all}}
\newcommand{\ffalse}{f_{false}^{\rm det}}
\newcommand{\find}{f_{ind}}

\begin{document}

\title[The reliability of the AIC method in Cosmological Model Selection] 
{The reliability of the AIC method in Cosmological Model Selection} 

\author[M.Y.J.~Tan and Rahul~Biswas]
  {M.Y.J.~Tan,$^{1}$
  Rahul~Biswas,$^{1,2}$\\ 
  $^{1}$Department of Physics, University of Illinois at Urbana-Champaign, Urbana, IL 61820, USA\\
  $^{2}$High Energy Physics Division, Argonne National Laboratory, Argonne, IL 60439-4815, USA}
\date{Released 2002 Xxxxx XX}

%

\maketitle 
\label{firstpage}

\begin{abstract} 
The Akaike Information Criterion (AIC) has been used as a statistical criterion to compare the appropriateness of different dark energy candidate models underlying a particular dataset. Under suitable conditions, AIC is an indirect estimate of the Kullback-Leibler divergence $D(T||A)$ of a candidate model $A$ with respect to the truth $T$. Thus, a dark energy model with a smaller AIC is ranked as a better model, since it has a smaller Kullback-Leibler discrepancy with $T$. 
In this paper, we explore the impact of statistical errors in estimating AIC during model comparison. Using a parametric bootstrap technique, we study the 
distribution of AIC differences between a set of candidate models due to different realizations of noise in the data and show
that the shape and spread of this distribution can be quite varied. 
We also study the rate of success of the AIC procedure for different values of a threshold parameter popularly used in the literature.  For plausible choices of true dark energy models, our studies suggest that investigating such distributions of AIC differences in addition to the threshold is useful in correctly interpreting comparisons of dark energy models using the AIC technique.
\end{abstract} 

\section{Introduction} 
Suppose we wish to choose the `best' model from a set of theoretical 
models (or theories) of a natural phenomenon with the aid of the relevant 
empirical data. How can we objectively accomplish this goal? 
Many statistical techniques such as hypothesis testing 
\citep{fisher..hypo}, Bayesian evidence \citep{Jeff61}, Akaike information criterion (AIC) \citep{1974ITAC...19..716A}, Bayesian Information Criterion (BIC) \citep{citeulike:90008}, minimum description length (MDL) \citep{citeulike:1738171}
, etc. have been developed to address the question of model selection.

The late time acceleration of expansion 
of the universe has been firmly established
\citep{1998AJ....116.1009R,1998ApJ...509...74G,1999ApJ...517..565P,2003ApJ...598..102K,2003ApJ...594....1T,2004ApJ...607..665R,2006A&A...447...31A,2007ApJ...666..694W,2009ApJ...700.1097H,2009ApJ...704.1036F,2009ApJS..185...32K,2010A&A...516A..63S,2011arXiv1105.2862B}, but there is no consensus on 
the physics behind this phenomenon. A number of possible explanations such
as a small positive cosmological constant or vacuum energy, an 
otherwise unobserved dynamical fields usually called dark energy
\citep{1988PhRvD..37.3406R,
1988NuPhB.302..668W,
1995PhRvL..75.2077F,
1999PhRvL..82..896Z},
or a modification of General Relativity
\citep{2000PhLB..485..208D,2001PhLB..502..199D,2005PhRvD..71f3513C} have been proposed as an explanation.
With many models still consistent with current 
data, it is clear that further progress of the field requires 
the collection of larger and complementary data sets and a definite 
framework for model selection. Several large new surveys such as 
\verb=DES=\footnote{\url{http://www.darkenergysurvey.org/}},
\verb=BIGBOSS= \footnote{\url{http://bigboss.lbl.gov/}} 
\verb=LSST= \footnote{\url{http://www.lsst.org/lsst/}} 
\verb=EUCLID=\footnote{\url{http://sci.esa.int/euclid}}
have been planned to study this late time acceleration by collecting 
more data \citep{Abbott:2005bi,2009arXiv0912.0201L,Schlegel:2009uw}.
Of course, even with accumulation of more quality data, the 
importance of analysing the model selection process will not diminish, 
because reliable discriminating methods can always allow us to exploit 
the available data maximally.

Hence, statistical techniques addressing model 
selection have been applied to this context 
\citep{Liddle:2004nh,Liddle:2006kn,2006PhLB..633..427S,2006PhLB..642..171S,Biesiada:2007um,Liddle:2006tc,2007ApJ...666..716D,2007MNRAS.377L..74L,2009ARNPS..59..95L,Trotta:2008qt,2009ApJ...703.1374S,Biesiada:2009zz,2011MNRAS.414.2337T,2011RAA....11..641B}. 
Many of these discuss the use of information criteria like AIC and BIC
which are easy to calculate. A number of works have applied this to data. In
particular, \citet{2006PhLB..642..171S} as well as \citet{Biesiada:2007um} used the method of AIC and BIC, and a compilation of SNIa to compare various late time acceleration cosmological models. 

\citet{2007ApJ...666..716D} and \citet{2009ApJ...703.1374S} compared such models based on the Sloan Digital Sky Survey \citep{2009ApJS..185...32K} and 
Equation of State: Supernovae Trace Cosmic Expansion(ESSENCE) Supernova data and high-redshift data \citep{2004ApJ...607..665R,2007ApJ...666..694W} along with a summary of cosmic microwave background (CMB) and Baryon Accoustic Oscillations (BAO) 
data using AIC and BIC. AIC has also been used in comparing 
principal component-based models for dark energy \citep{2007PhLB..648....8Z}. 
We wish to explore the use of AIC in the context of 
selecting the `best' theoretical model describing the late time acceleration
 of the universe.

This paper will focus on the use and validity of applying the AIC technique 
to choose the `best' late time acceleration model from SNIa data. 
For our calculation, we use data from the Constitution compilation 
\citep{2009ApJ...700.1097H} of the Center for Astrophysics (CFA3) sample 
\citep{2009ApJ...700..331H}
ESSENCE \citep{2007ApJ...666..674M}, 
Supernova Legacy Survey (SNLS) \citep{2006A&A...447...31A} and 
`High-$z$' samples \citep{2007ApJ...659...98R}. 
We will study and refine the AIC methodology in this context. 
In particular, we wish to study its reliability when subjected to estimator uncertainty. 
With the AIC technique, the usual approach has been to use a single number (called AIC) to rank models; the smaller the AIC the better the model. 
This concept of discrete ranking of models has been extended \citep{Jeremy_cite:1,citeulike:11258} by paying attention to the differences between actual AIC values. 
However, it is important to note that the AIC values themselves are 
empirically estimated from the data and thus have statistical uncertainties. Therefore, the reliability of the estimates of AIC differences is a crucial issue. 
While there is work \citep{RePEc:spr:aistmt:v:49:y:1997:i:3:p:395-410} that tries to address the reliability of the AIC technique through 
analytic calculations, we will instead study the reliability through 
numerical simulations.

The paper is organised as follows. In Section \ref{AIC}, we briefly review the information 
theoretic origin of AIC, extending this idea further, by reviewing the theoretical implications of the AIC differences in Section \ref{Model_Comparison}.
In Section \ref{sec:reliability}, we study the estimator uncertainty of these AIC differences in the context of cosmological model selection. 
This involves a comparison of a set of four candidate models by using SNIa, as in the work by \citet{2006PhLB..633..427S}. 
The reliability of these AIC estimates are calculated via a bootstrap method \citep{citeulike:1616452}, and this will be used to evaluate the validity of the model selection technique. 
The results of these simulations and the conclusion will then be given in Sections \ref{boot_strap} and \ref{conclusion}, respectively.
\section{The Akaike Information Criterion (AIC)}
\label{AIC}
The Kullback-Leibler (KL) divergence \citep{citeulike:165404} is a commonly 
used quantity that measures the discrepancy \footnote{Note that a KL divergence is a non-symmetric measure that does not obey the triangle inequality.} of one probability distribution with respect to another probability distribution.
We denote the KL divergence of a candidate model $A$ (which gives the probability distribution $p_A$) with respect to the truth $T$ (with probability distribution $p_T$) as $D(T||A)$, where we have defined $T$ as an underlying process consisting of a signal with stochastic noise; a particular empirical datum is a single realisation of this process. If we denote $X$ as the set of all possible outcomes that can be generated by 
either $A$ or $T$, and $x$ as an element in this set, we can define $D(T||A)$ as
\[
 D(T||A) = \int_{x \in X} dx \ p_T(x) \log\frac{p_T(x)}{p_A(x)}.
 \] 
In this paper, we define a model class as the totality of model probability distributions with the same parametric form (but with different parameter values). Within each model class, there is a set of parameters (the `best' model) that gives the lowest KL divergence with respect to the truth. Thus, to choose the `best' model class we must first choose the `best' model (parameter set) from a particular model class as the representative of the model class. This is done by the maximum likelihood criterion. The model class selection strategy is thus obtained by comparing the KL divergence of the representative models of the individual model classes.\\ 
However, the truth is unknown a priori, so for a given representative model $A$, $D(T||A)$ cannot be evaluated directly. We can solve this problem by computing the AIC value \citep{1974ITAC...19..716A} which is an asymptotically 
unbiased estimator for $D(T||A)$ (up to a fixed offset that is independent of the representative models). Since the fixed offset is independent of the models, and hence the choice of model classes, a comparison of the AIC values is a useful surrogate for the strategy of comparing the associated KL divergence of the different candidate model classes. We can compute the AIC of the representative models of various model classes, which in the asymptotic limit is known \citep{1974ITAC...19..716A} to be 
\begin{equation} 
\mbox{AIC} = 2k - 2\log(\mbox{L}_{ML}),
\label{eqn:AIC_0}
\end{equation}
where $\mbox{L}_{ML}$ denotes the likelihood evaluated at the set of 
model parameters that maximize the likelihood $\mbox{L}$, and $k$ is the number of free 
parameters in the candidate model class. When we make a further assumption that the distribution of errors follows a Gaussian distribution, this further reduces to
\begin{equation} 
\mbox{AIC} = 2k + \chi^2_{ML},
\label{eqn:AIC}
\end{equation}
where $\chi^2_{ML}$ is the usual chi-square evaluated at the maximum 
likelihood estimate of the model parameters. The AIC values are subsequently ranked by the smallness of their values; the model class with the smallest value is determined to be the `best' model class.

It is worth noting that the AIC estimate (Eqn.~\ref{eqn:AIC}) of the KL divergence assumes that the number of data points is sufficiently large. 
AIC in the form written above is an unbiased estimator for large 
data sets. For smaller data sets, the $2k$ term can be corrected by an additional $\frac{2k(2k+1)}{N-k-1}$ 
term to approximately correct for the bias due to finiteness of the dataset, where $N$ is the number of data points in a single data set. 
While further studies to obtain a more accurate expression for this term 
are possible, for the cases we shall consider, this correction is always less than $0.06$ 
(which will be seen to be negligible for our purposes) and will only decrease in importance when 
more data is collected. We shall therefore ignore this correction altogether in this paper.

As already mentioned explicitly, the AIC procedure for comparing different model classes actually compares
the $\chi^2$ for the best fit representative model from each model class. These representative models are derived from the maximum
likelihood estimate of the model parameters for each of the model class. Thus, in the rest of this paper, when we refer to comparing model classes, the calculation only involves the $\chi^2$ of the (best) representative model taken from its respective model class. We sometimes refer to this as comparing models.

\section{Model Comparison and AIC differences}
\label{Model_Comparison}
Since AIC is essentially the measure of discrepancy of a model from the truth, it is intuitively obvious that the smaller the AIC difference 
between two models, the harder it becomes to judge which model is better; even if the AIC estimate of this difference in the KL divergence can be obtained without an estimator error, the small difference would make it difficult to tell the two probabilistic models apart for a small number of observations. Hence, there is a need to associate a confidence level for distinguishing between a model $A$ and another model $B$ using the AIC difference between them. 

Let $P(M)$ be the probability that model $M$ is true. We wish to relate $P(A)/P(B)$ to the AIC difference 
\[
\Delta_{A,B} \equiv \mbox{AIC}(A)-\mbox{AIC}(B) 
\approx 2\left[D(T\vert\vert A)-D(T\vert\vert B)\right], 
\]
 
where $\approx$ implies the asymptotic relation. This asymptotic
relation may not necessarily be realised as the number of
samples is small. Using a certain set of extra assumptions,
\citet{Jeremy_cite:1} showed that for a pair of models $A$ and $B$,
\begin{equation}
P(A)/P(B) \approx 
\exp[-\Delta_{A,B}/2].
\label{odds}
\end{equation}
Since the AIC differences are estimates of the differences of KL divergences, one can also use the idea of distinguishing probability 
distributions to justify Eqn.~\ref{odds} when the number of samples are sufficiently large (Appendix~\ref{app:Confusion}). Assuming the truth to be in the set of candidate models, \citet{citeulike:11258} extended Eqn.~\ref{odds} to obtain the probability $w_i$ of model $i$ by appropriately normalising 
the equation:
\begin{equation} 
w_i=\frac{\exp[-\Delta_{i,b}/2]}{\sum_j^n \exp[-\Delta_{j,b}/2]}, 
\label{modellikelihood}
\end{equation} 
where the candidate models are numbered as $ i = 1, \cdots ,n$ and $b$ denotes the best model which has the smallest AIC value among all the candidate models. 
Either way, Eqn.~\ref{odds} quantifies the intuitive idea that it is easier
to select a model over another if the AIC difference is large. 

One can use this idea to modify the AIC methodology described above 
and suppress the probability of obtaining incorrect results by introducing 
a threshold $\dt$. 
Then, rather than ranking all models according to the smallness of their 
AIC values, one adopts the procedure where a model A is ranked to be better 
 than model B if  $\Delta_{A,B}
< -\dt$, while any two models with an AIC 
difference smaller than $\dt$ are  considered of equal rank. 
Eqn.~\ref{odds} shows that choosing a large  enough  value of the 
threshold $\dt$ implies a high probability that the selected model is 
truly the better one. However, a large value of $\dt$ also increases the 
number of model pairs where the AIC differences  are in the range 
$-\dt < \Delta_{A,B} < \dt.$ Since this procedure cannot 
discriminate between such models,  we shall call such a model selection 
result indeterminate. In our convention, we also define the converse of the indeterminate case as the determinate case ($\vert \Delta_{A,B}\vert > \dt$). 
Of course, for 
a  pre-determined choice of $\dt$ (corresponding to a predetermined 
confidence level), a better dataset gives a smaller fraction of model pairs 
which have indeterminate results.  
As a rule of thumb, a universal value of the threshold $\dt = 5$, without any regard 
to the properties of the models under comparison, has been mentioned by 
\citet{2007MNRAS.377L..74L} as the minimum AIC difference 
between two models needed to make a `strong' assertion that one model is 
better than the other. Such a definition has been used extensively in the 
literature.
\section{Impact of AIC uncertainties in finite data sets}
\label{sec:reliability}

In the preceding section, we have discussed the AIC differences $\Delta_{A,B}$ without any regard for the fact that AIC is a statistical estimate. The associated uncertainty in the AIC estimate may not be negligible and may be dependent on the realisation of noise in a particular 
data set. Thus, there must be a statistical uncertainty in the value of $\Delta_{A,B}$
\footnote{
This is due to a statistical uncertainty in the AIC values coming from 
the variation of $\chi^2_{ML}$. However, the uncertainty in the AIC values of the models
can be correlated, and turns out to be larger than the uncertainty in the 
AIC differences.}
even when estimated from a data set of similar quality
\footnote{In this paper, two Supernovae data sets are said to have the same quality when they have the same number of data points, the same set
of redshift $z$ values and the same set of error bars(standard deviation) that corresponds to the set of $z$ values.}. 
The ensemble of such observations defines an empirical probability distribution of $\Delta_{A,B}$, and the particular 
value of $\Delta_{A,B}$ obtained from the current SN data sets is actually a sample value drawn from this probability distribution. 
\\
\\
Ideally, we should be able to study the probability density distribution of $\Delta_{A,B}$ under repeated observations of results with sample size $N$: $P(\Delta_{A,B}\vert E_N)$, where $E_N$ denotes a collection of observation data which individually consists of $N$ observation points and are drawn from the underlying truth process. Because producing a large subset of $E_N$ is impossible, in this paper we instead use a bootstrap approach \citep{citeulike:1616452} to generate `mock' empirical data sets and estimate the probability distribution of $\Delta_{A,B}$.
\\
\\
Perhaps the most frequently used method to produce `mock' empirical data is the bootstrap method proposed by \citet{citeulike:1616452}. When we apply this approach to regression models, we need a probability model that specifies the distribution of residuals (eg. Gaussian distribution). In our case in particular, the distance modulus $\mu$ is related to the red shift $z$, so a regression model has the following structure $\mu = f(z) + \epsilon$, where $\epsilon$ is the residual (the error term). Since we do not know the true relation $f$, we cannot obtain the purely empirical distribution of the residuals
\footnote{Note that these errors which include light curve fitting errors, intrinsic dispersion and peculiar velocity corrections are also required for calculating quantities like $\chi^2$ for most model selection schemes. Thus, an underlying assumption of the application of the AIC technique as in previous works is that these error estimates are correct. Since our focus is on the statistical uncertainties in AIC after following other underlying assumptions used in the literature, we also assume that these error estimates are 
correct.}. Thus, for regression models, the standard bootstrap method always involves using a model-dependent probability distribution of residuals. \citet{2007ApJ...666..716D} extended this method to compare between two regression models and find the standard deviation of their BIC differences; we will further extend this idea and show that studying the structure of the distribution of AIC differences is important.
\\
\\
When we wish to check the reliability of a particular model, we could use the probability distribution of residuals based on the model itself. Since we cannot have any model-free bootstrap data, we must choose a particular model to produce the residual distribution. Therefore, to estimate $\Delta_{A,B}$, we need some model $C$ as a reference probability model that is used to generate the bootstrap data. Let us denote the estimate of $\D_{A,B}$ based on a data set $d$ as $\D^d_{A,B}$. We wish to produce $\{\Delta^{d}_{A,B}\vert d\in C_N\}$, where $C_N$ denotes a collection of parametric bootstrap data generated by model $C$, which individually consist of $N$ observation points with the same data quality as our empirical data. Although the needed details are in Appendix~\ref{app:Bootstrap}, our approach may be outlined as follows. 
\\
\\
Suppose we have a single data set consisting of $N$ observed results $\{(z_1,\mu_1,\s_1),(z_2,\mu_2,\s_2),\cdots ,(z_N,\mu_N,\s_N) \}$, 
where the $z_i$ values denote the $i^{th}$ observed redshift, $\mu_i$ the $i^{th}$ generated distance modulus, and 
$\s_i$ the observed error bars of $\mu_i$. We can create a bootstrap sample $C_N$ consisting of $N$ observation points based on model $C$. $C_N$ relates the same set of coordinates by the relation: $\mu_i = f(z_i) + \e_i$, $i=1, \cdots, N$, where $\e_i$ is 
a stochastic term obeying a normal distribution of mean 0 and standard deviation $\s_i^2$: $N(0,\s_i^2)$, and $f(z)$ is the maximum likelihood estimate of model $C$ ($\Lambda$CDM, DGP, etc.) (Appendix~\ref{app:hubble}). 
\\
\\
As mentioned above, we wish to simulate the distribution of $\{\Delta^{d}_{A,B}\vert d\in C_N\}$ as a proxy for $P(\Delta_{A,B}\vert E_N)$. To proceed, we choose a subset of models that have been often studied in the literature \citep{2006PhLB..633..427S} and are listed in Table.~\ref{tab:candidatemodels}. We also assumed that the universe is spatially flat and set the curvature term $\Omega_{k}$ in the Hubble function $H(z)$ to zero. Three of these models $\Lambda$CDM, wCDM, CPL \citep{2001IJMPD..10..213C,2003PhRvL..90i1301L}
are dark energy models with different parametrisations of the equations of state $w(z)=p(z)/\rho(z)$, where $p(z)$ and $\rho(z)$ 
are the pressure and density of dark energy, respectively. 
These models are nested: setting $w_a=0 $ in the CPL model, we obtain the wCDM model; setting $w_0=-1$ in the latter gives the $\Lambda$CDM model. 
We also use the flat DGP model which is a modified 
gravity model and cannot be nested in the previous classes of models. 
\\
\\
To make contact with observational data, we choose candidate models with parameter sets that are `best' for the Constitution
compilation of SNIa data ($f(z)$ being the maximum likelihood estimation derived model), 
where we use the distance moduli and the error bars in data (Appendix~\ref{app:hubble}). Since 
we need to find estimates of cosmological parameters for different 
models by maximising likelihoods, we use the results from the 
more appropriate MLCS light curve fitter (for $R_V = 1.7$) in \citep{2009ApJ...700.1097H}. 
The details of finding the maximum likelihood and the corresponding AIC value is given in Appendix~\ref{app:hubble}; 
371 SNIa events were used. The Hubble function, as a function of cosmological parameters in each of 
these models, along with the free parameters and AIC values (calculated 
from the Constitution compilation) are shown in Table.~\ref{tab:candidatemodels}.
\begin{table*} 
\caption{Different model classes considered in this paper: We show the evolution of the Hubble function $H(z)$ with
redshift $z$, the Hubble constant $H_0$ and other free parameters in the models. $k$ is the number of free parameters. The respective AIC values were obtained from the Constitution compilation.} 
\label{tab:candidatemodels} 
\begin{tabular}{|c|c|c|c|c|} 
\hline 
Model & $H(z)/H_0$ & Free Parameters & $k$ & AIC \\ 
\hline 
$\Lambda$CDM  & $\sqrt{\Omega_{m}(1+z)^3+ (1-\Omega_{m})}$ & $\Omega_{m}$ & 1 & 401.35 \\ 
\hline
wCDM  & 
$\sqrt{\Omega_{m}(1+z)^3+ (1-\Omega_{m})(1+z)^{3(1+w)}}$ & 
$\Omega_{m}$,$w$ & 
2 & 
403.05 \\ 
\hline
CPL  & 
$\sqrt{\Omega_{m}(1+z)^3+ (1-\Omega_{m})(1+z)^{3(w_{0}+w_{a}+1)} \exp[\frac{-3w_{a}z}{1+z}]}$ & 
$\Omega_{m}$,$w_0$,$w_a$ & 
3 & 
404.66 \\
\hline 
DGP  & 
$(\sqrt{\Omega_{m}(1+z)^3+ \Omega_{rc}} +\sqrt{\Omega_{rc}})$ &
 $\Omega_{m}$ & 
1 & 
401.13 \\ 
  &
$\Omega_{rc}=(1-\Omega_{m})^2/4$ &
 &
 & \\ 
\hline 
\end{tabular} 
\label{model_list}
\end{table*} 
\\
\\
For the nested models, the parameter values that fit the data best turn out 
to be close to the $\Lambda$CDM model. Consequently, one does not
gain much in terms of a lower $\chi^2_{ML}$, while the extra free parameters are 
penalised to give higher values of AIC for wCDM and CPL.
The DGP model gives the best AIC value, which is only slightly better than the 
$\Lambda$CDM model. It is known that the simultaneous use of 
Cosmic Microwave Background (CMB) data from WMAP \citep{2007ApJ...666..716D,2009ApJ...703.1374S} and Large Scale Structure (LSS) data 
disfavours the DGP model compared to $\Lambda$CDM, since the 
parameter subspaces that provide the best fits for CMB, LSS 
and SNIa data do not overlap as much as in the $\Lambda$CDM model. 
We checked that our results are consistent with this, but will ignore the CMB
and LSS data to focus on methodology.
\\
\\
In order to calculate the distribution of $\Delta^d_{A,B}$, we adopt the best fit model in a model class $C$ as a reference model to produce 5000 mock data sets of $C_{371}$ which is expected to be similar in quality to the Constitution compilation of Supernova data. All the model classes in Table.~\ref{tab:candidatemodels} are successively chosen as the reference model $C$. Following our definition of `similar quality', each simulated data set has the same set of redshift values and error bars as the Constitution compilation, while their apparent magnitudes are those expected from
a noisy realisation of the reference model. The basic steps are:
\begin{enumerate} 
\item We produce a mock data set consisting of 5000 realisations of $d \in C_{371}$ for a reference model $C$ as outlined above.
\item Candidate models $A$ and $B$ are fitted to $d \in C_{371}$ by maximising the likelihood and the AIC values of these models $A$ and $B$ are computed through Eqn.~\ref{eqn:AIC}.
\item Thus, for each element $d \in C_{371}$ we can make the AIC difference $\D^d_{A,B}$. We study $\{ \Delta^d_{A,B} \ \vert \ d\in C_{371}\}$ for the reference model $C$ by plotting a histogram.
\end {enumerate} 
We should note that the probability distribution of $\Delta^d_{A,B}$
is due to errors introduced by the stochastic noise term, but ignores the effect of the uncertainties in 
the cosmological parameters of the reference model $C$. 

\section{Bootstrap Study Results}
\label{boot_strap} 
\begin{figure} 
\includegraphics[width=60mm]{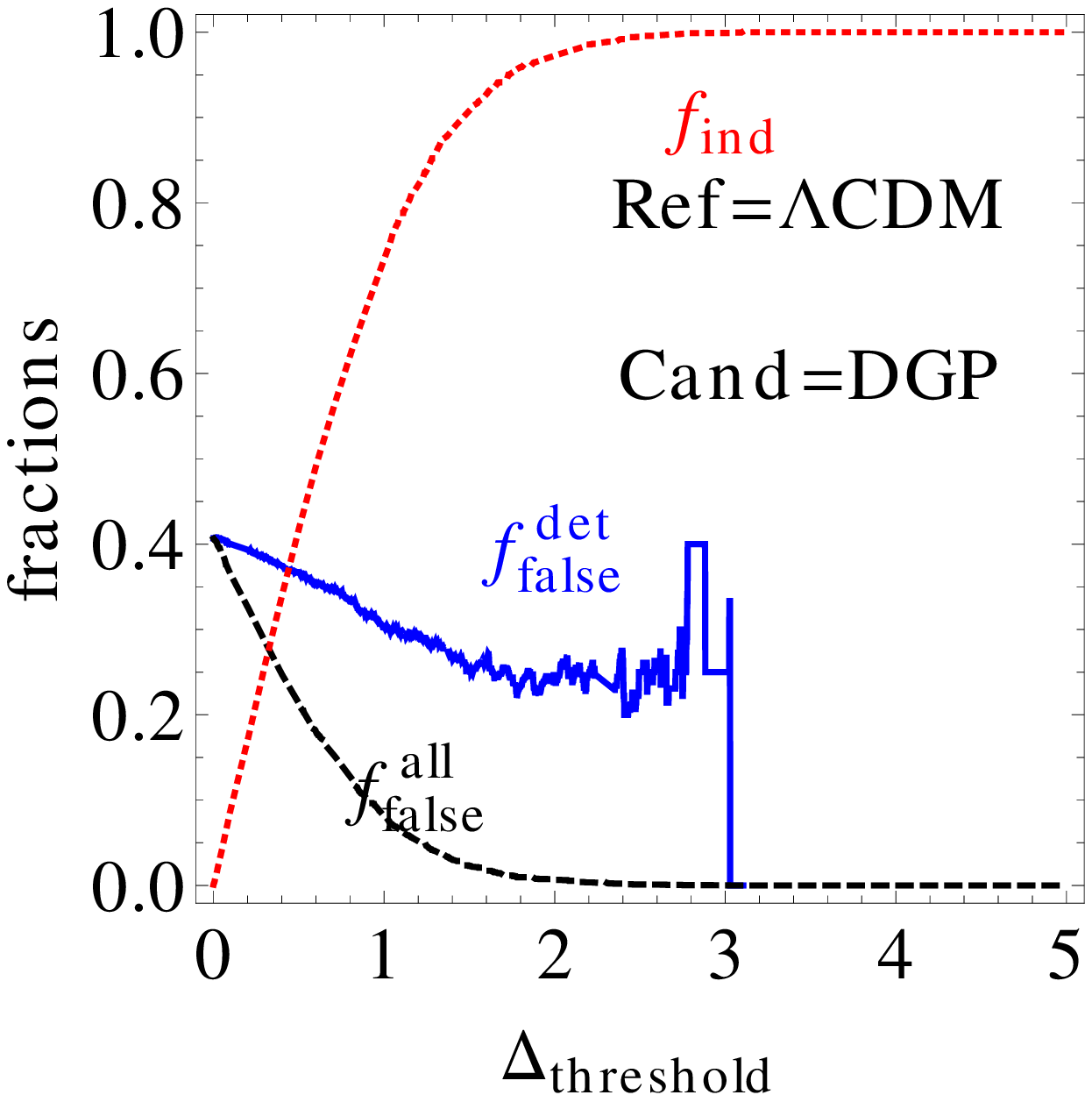}\\
\includegraphics[width=60mm]{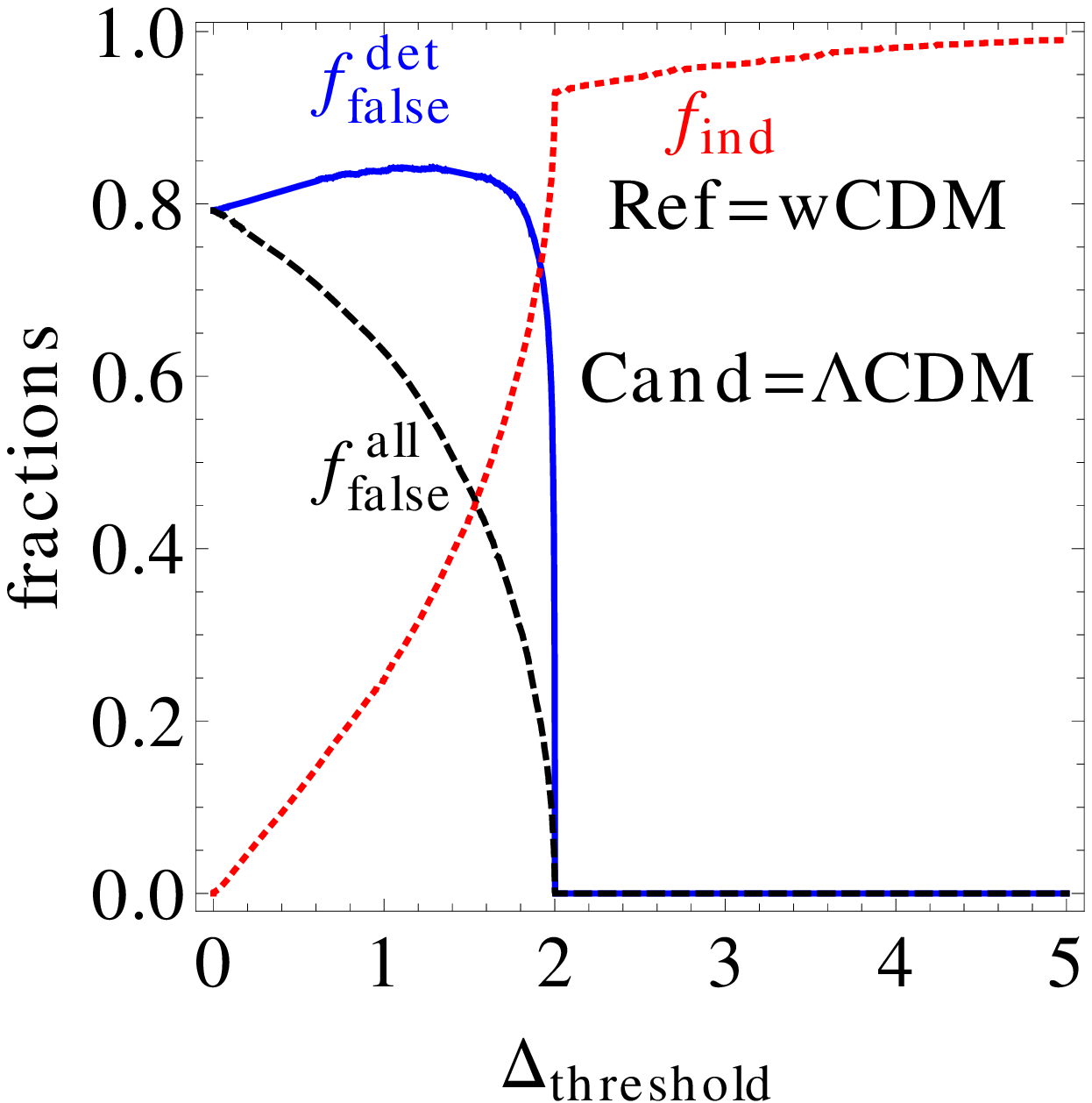}\\
\includegraphics[width=60mm]{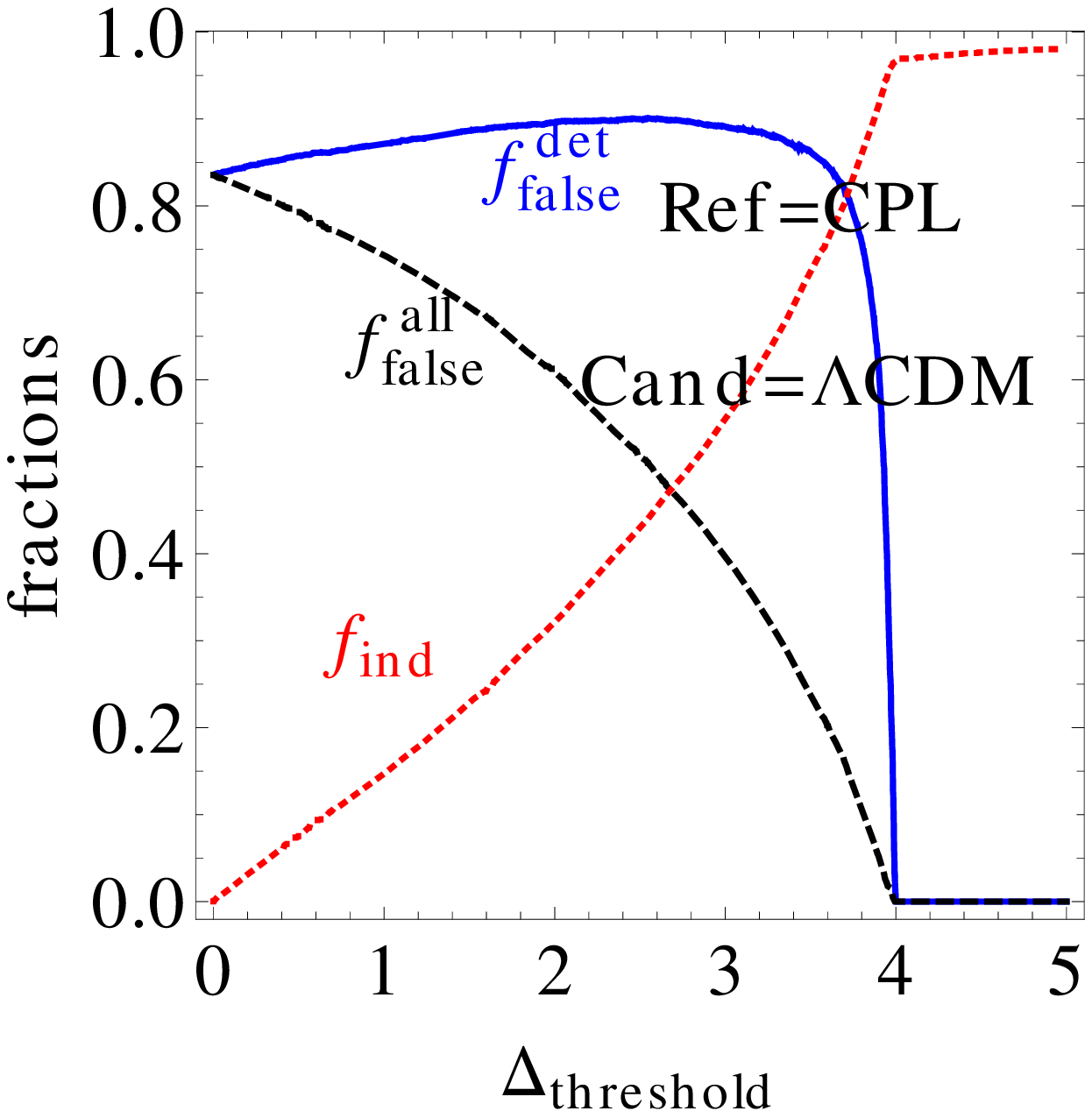}
\caption{~Probabilities of different candidate models (Cand) 
being selected over different reference models (Ref) by using AIC 
for different values of $\dt$ for the case of 
(a) (top) 
the candidate model DGP being picked over the reference model $\Lambda$CDM
(b) (middle) the candidate model $\Lambda$CDM being picked over the reference model wCDM, and 
(c) (bottom) the candidate model $\Lambda$CDM being picked over the reference 
model CPL. 
The solid 
blue curve shows the number of incorrect results as a fraction
$\ffalse$ of the cases where the procedure returns a determinate
result. The black dashed curve shows the number of incorrect results as a
fraction of the total number of simulations. 
The fraction $\ffalse$  is extremely noisy and should be ignored when the
fraction of determinate cases $\find$ (red, dotted) is large. The plots 
show that increasing $\dt$ always decreases $\nf$, $\ffalse$ does not necessarily decrease. For a comparison of the models used, this shows that AIC tends to incorrectly 
select models with a lower number of parameters. In the comparison of nested models,
 $\ffalse$ drops sharply at twice the difference in the number of free parameters of the 
models compared.}
\label{FailureRatevsthreshold} 
\end{figure}

\begin{figure} 
\includegraphics[width=60mm]{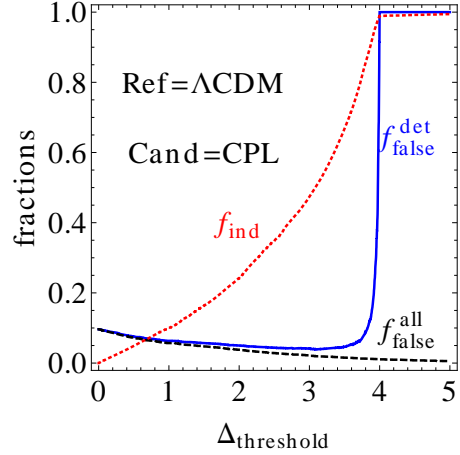}
\caption{~Probability of the candidate model (Cand) CPL being selected over the reference model (Ref) $\Lambda$CDM for different values of 
the threshold. In this case, $\ffalse$ actually increases sharply at about
twice the difference in free parameters between these models, showing that 
in this case AIC tends to incorrectly select the model with a higher number of 
parameters. } 
\label{FailureRatevsthreshold-diff} 
\end{figure}

We first consider the issue of statistical self-consistency in the following sense: When the reference model is $C$, does the bootstrap AIC method outlined above choose model $C$ as a better model than the rest? The distribution of the values of $\Delta^d_{C,A}$ for $d \in C_N$ can tell us about the statistical self-consistency. We start with the case when $\Delta_{threshold}=0$. If $P(A/C) \equiv P(\{d \in C_N\vert \Delta^d_{C,A}<0\})$ is larger than some predefined proportion, which must not be less than 1/2, we may say model $C$ is better than $A$ when $C$ is the reference. The $P(A/C)$ results for this case is summarised in Table.~\ref{signn}. The P-values \footnote{The highest P-value found was $3 \times 10^{-8}$ for $\Delta_{DGP,\Lambda CDM}$. Most of the P-values found were at the level of machine precision.} for the table are extremely small.
\begin{table} 
\begin{center} 
\begin{tabular}{cccccc} 
\hline 
\hline 
X$\setminus$Y & DGP & $\Lambda$CDM & wCDM & CPL \\ 
\hline 
\\[0.05ex] 
DGP  & - & 58 $\%$ & 84 $\%$  & 89$\%$  & \\ 
 & & & & \\ 
$\Lambda$CDM & 59$\%$ & - & 85$\%$  & 90$\%$  & \\ 
 & & & \\ 
wCDM  & 16$\%$ & 21$\%$ & -  & 90$\%$  &  \\ 
 & & &  \\ 
CPL  & 14.0$\%$ & 16$\%$ & 17$\%$  & -  & \\ 
\\[0.1ex] 
\hline 
\end{tabular} 
\caption{
Percentage of cases where the AIC method with a threshold $\dt=0$ selects the correct model over 
other candidate models considered. We defined the difference and percentage using the following convention. If we define the bootstrap data as being produced by the reference model $X$ (rows) and we are comparing it against model $Y$ (columns), the difference is defined as the AIC value of $X$ minus the AIC value of $Y$ and is denoted by the symbol $\Delta^d_{X, Y}$. The table counts the percentage of negative $\Delta^d_{X, Y}, d\in X_N$. Note that a value greater than 50$\%$ indicates that the correct model is chosen a majority of the time. We use this cut-off of 50$\%$ or more as a simple definition for statistical self-consistency.
}\label{signn}
\end{center}
\end{table}  
\\
\\
From Table.~\ref{signn}, we obtained $\Delta^d_{DGP,CPL}<0$, 89$\%$ of the time, when DGP model is the reference (i.e., $P($CPL/DGP$) = 0.892$), and 83$\%$ of the time, when CPL model is the reference (i.e., $1-P($DGP/CPL$)) = 0.836$). This means that, when we use a threshold $\dt=0$, the DGP model is significantly favoured over the CPL model even if the CPL model is the reference. This means that the AIC method is statistically inconsistent when it is applied to the CPL model, and would be automatically disqualified under the current number of data points in the sample. In another example, we compared the DGP and $\Lambda$CDM models in the table and noticed $\Delta^d_{\Lambda CDM,DGP}<0$, 59$\%$ of the time when $\Lambda$CDM model is the reference (i.e., $P$(DGP/$\Lambda$CDM) = 0.594), and 42$\%$ of the time when DGP model is the reference (i.e., $1-P(\Lambda$CDM/DGP) = 0.422). This means that if the reference model was either $\Lambda$CDM or DGP and we apply AIC to compare between them using a zero $\dt$, AIC will only slightly favour the reference. 
The AIC technique is only statistically self-consistent (for a zero $\dt$) when applied to compare DGP and $\Lambda$CDM while we cannot use AIC to self-consistently study the other models under the current level of observation quality. However, a test that gives the right answer 3 out of 5 times is unreliable, since we can only do a single empirical test from our actual data. $\Lambda$CDM and DGP cannot be distinguished significantly using either reference models. Thus, looking at both examples, we must conclude that there are insufficient data points to tell reliably the models apart using AIC when $\Delta_{threshold}=0$. Another trend that results from the insufficiency of data points is that the AIC procedure tends to favour models with a smaller number of free parameters. The trend persists even when we later increase $\dt$. This seems to indicate that the addition of extra free parameters does not significantly improve the $\chi^2$ fit for the number of data points used.
\\
\\
We next study the behaviour when the threshold 
parameter is increased, $\dt = 2$ and $5$, corresponding to choices made in the literature to moderate and strong evidence. Unlike the previous $\dt = 0$ case, we have to consider the effect of indeterminate cases, which we defined in section~\ref{Model_Comparison} as the case when $\vert \Delta_{A,B}\vert < \dt$.
\\
\\   
In order to study the reliability of the AIC technique at 
different value of the threshold parameter $\dt>0$, we
analyse the probability of the selected model being incorrect 
for different values of $\dt$. To do so, we define the following:
\\
\begin{eqnarray}
\find &=& 
\frac{\mbox{Number of samples with } \vert \Delta_{A,B}\vert < \dt}{\mbox{Number of samples}}\\
\nf &=& \frac{\mbox{Number of samples with } \Delta_{A,B} > \dt}{\mbox{Number of samples}}\\
\ffalse &=& 
\frac{\mbox{Number of samples with } \Delta_{A,B} > \dt}{\mbox{Number of samples}
(1-\find)
} 
\end{eqnarray}
$\find$ is the 
fraction of cases where the AIC procedure has an indeterminate result for a given value of the threshold $\dt$, so a high 
value of $\find$ reflects the inadequacy of the data to discriminate 
between the pair of models in question with a certain level of confidence for a relevant $\dt$.
Using $A$ as the reference model, $\nf$ is the fraction among all cases, where the AIC procedure results in 
an incorrect model selection. 
$\ffalse$ is the fraction among determinate cases ($\vert \Delta_{A,B}\vert > \dt$) where the AIC 
procedure results in an incorrect selection, and reflects the ratio of 
correct to incorrect model selections. Our results are summarized in Tables.~\ref{Tab:failurerate-2} and~\ref{Tab:failurerate-5}. In each 
table, the rows correspond to different reference models $X$, while the 
columns list  candidate models $Y$ that were compared with the 
reference model.
\\
\\ 
For each pair of reference model $X$ and candidate model $Y$, we show the 
fractions $(\find, \nf , \ffalse)$. 
First we reconsider the $\dt=0$ case in Table.~\ref{signn}. 
By definition, the proportion of $\dt=0$ cases where the AIC method is not statistically self-consistent ($1-P(A/B)$) is equivalent to both $\nf$ and $\ffalse $.
We note that the values of $f_{false}$ are large, indicating a high failure rate and an unsatisfactory procedure.
\\
\\
We expect these failures to be suppressed when we choose higher values of the threshold $\dt$. When we increase the threshold $\dt$ to 2 and then 5, we notice the expected suppression of $\ffalse$. However, $\ffalse$ does not decrease as dramatically as $\nf$.
\\
\\
We study the behaviour of $\ffalse,$ $\nf$ and $\find$ in Fig.~\ref{FailureRatevsthreshold} for different values of $\dt$ as well as 
different choices of candidate and reference models; using bootstrap simulations. 
 The $\ffalse$ values become dominated by noise as $\find$ increases, 
since the calculation is made from an ever decreasing number of determinate bootstrap cases. Hence, the values of $\ffalse$ 
 at large values of $\find$ should be ignored. Nevertheless, we can still 
study the regime for smaller $\find$. It should also be noted that for the case of $\Delta_{threshold}=5$, the proportion of indeterminate results was high (Table.~\ref{Tab:failurerate-5}). For example, when CPL is the reference model, the proportion of $\Delta^d_{CPL,\Lambda CDM}$ and $\Delta^d_{CPL,DGP}$ between $\pm 5$ are both approximately 98$\%$.
\\
\\
We also note, as expected, that $\find$ increases asymptotically to one as we increase $\dt$. 
Increasing $\dt$ monotonically suppresses $\nf$, but not necessarily 
$\ffalse$. $\ffalse$ decreases with increasing $\dt$ for the cases where 
the DGP model was wrongly picked over the reference $\Lambda$CDM model. However,
for the other cases, when the candidate model has a smaller number of 
free parameters than the reference model, $\ffalse$ tends to gently 
increase before steeply decreasing at a certain value of $\dt$. From Fig.~\ref{FailureRatevsthreshold}, 
we can see that this sharp decline happens at roughly twice the 
difference in the number of free parameters between the candidate and reference models. We also study these quantities for the case 
where the reference model has less parameters than a candidate model in 
Fig.~\ref{FailureRatevsthreshold-diff}. In this case, $\ffalse$ actually 
increases rapidly at the point where $\dt$ is equal to the 
difference in the number of free parameters in the two models. This implies 
that if the model underlying the empirical data was similar to the reference models studied, it is improbable that the data 
set would provide an AIC difference for the considered models greater than
a large threshold $\dt$ (eg. 5) as shown by $\find$. However, if this dataset 
did yield a AIC difference larger than a predetermined $\dt$, it does 
not necessarily mean that the AIC selected model has a high probability of 
being the true underlying model. This is because $\ffalse$ for a given $\dt$ depends on the model pairs
 being considered, suggesting that even having AIC differences 
larger than a specified threshold $\dt$ does not guarantee reliability of 
the AIC selection process. 
\\


\begin{table*} 
\begin{center} 
\begin{tabular}{cccccc} 
\hline 
\hline 
X$\setminus$Y & DGP & $\Lambda$CDM & wCDM & CPL \\ 
\hline 
\\[0.05ex] 
DGP  & - & (97,1,26)$\%$  &(86,5,32)$\%$  & (25,4,5)$\%$& \\ 
 & & & & \\ 
$\Lambda$CDM & (97,1,24)$\%$ & - & (94,4,66)$\%$  & (24,4,5)$\%$  & \\ 
 & & & \\ 
wCDM  & (87,9,71)$\%$ & (90,1,11)$\%$ & -  & (96,3,79)$\%$  &  \\ 
 & & &  \\ 
CPL  & (31,64,92)$\%$ & (32,61,90)$\%$ & (94,0,0)$\%$  & -  & \\ 
\\[0.1ex] 
\hline 
\end{tabular} 
\caption{Failure Rate for $\dt=2:$ The values in the parentheses show 
$\find$, the percentage of total cases where the AIC procedure has an 
indeterminate result; $\nf$, the fraction of total cases where the AIC 
procedure results in an incorrect model selection; and $\ffalse$, the percentage of determinate cases where the AIC wrongly selects a candidate model Y over the reference model X.}
\label{Tab:failurerate-2}
\end{center}
\end{table*} 
\begin{table*} 
\begin{center} 
\begin{tabular}{cccccc} 
\hline 
\hline 
X$\setminus$Y & DGP & $\Lambda$CDM & wCDM & CPL \\ 
\hline 
\\[0.05ex] 
DGP  & - & (100,0,``-")$\%$ & (99,1,100)$\%$  & (99,1,100)$\%$  & \\ 
 & & & & \\ 
$\Lambda$CDM & (100,0,``-")$\%$ & - & (99,1,100 )$\%$  & (99,1,100)$\%$  & \\ 
 & & & \\ 
wCDM  & (100,0,0)$\%$ & (99,1,0)$\%$ & -  & (99,1,0)$\%$  &  \\ 
 & & &  \\ 
CPL  & (99,0,0)$\%$ & (98,0,0)$\%$ & (99,0,0)$\%$  & -  & \\ 
\\[0.1ex] 
\hline 
\end{tabular} 
\caption{Failure Rate for $\dt=5:$ The values in the parentheses show 
$\find$, the percentage of total cases where the AIC procedure has an 
indeterminate result; $\nf$, the fraction of total cases where the AIC 
procedure results in an incorrect model selection; and $\ffalse$, the percentage of determinate cases where the AIC wrongly selects a candidate model Y over the reference model X.}
\label{Tab:failurerate-5}
\end{center}
\end{table*} 

\begin{figure} 
\includegraphics[width=84mm, height=192mm]{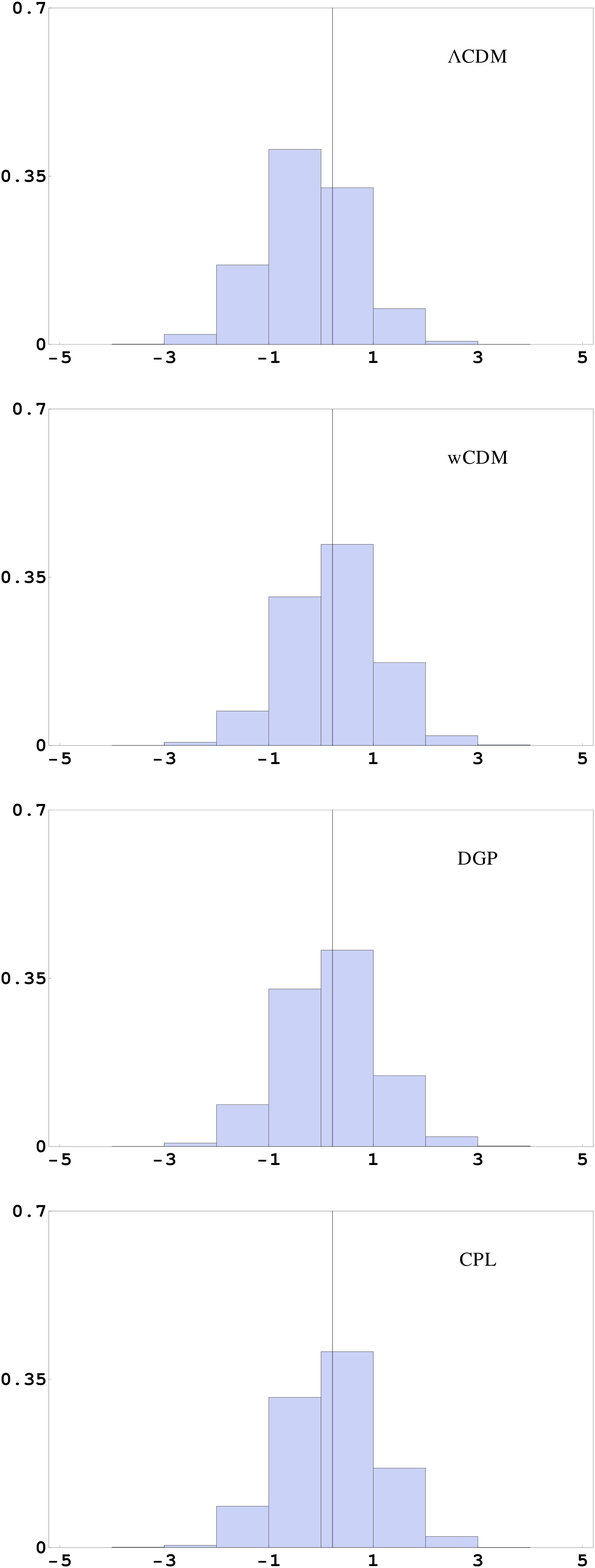}
\label{DGPLCDMcomparison1}
\caption{~Probability distributions of the AIC differences between $\Lambda$CDM and DGP ($\Delta^d_{\Lambda CDM,DGP}$) for different reference model $C$. $C$ is used to generate the respective bootstrap samples and is written in the right upper corner of the figures. The horizontal axis indicates the $\Delta^d_{\Lambda CDM,DGP}$ value while the vertical axis indicates their relative frequency. If the process underlying our observations was really the best fit model of class $C$, then the values of the AIC differences under different realisations of noise would have the histogram distribution $\{\Delta^d_{\Lambda CDM,DGP} \ \vert \ d\in C_{371}\}$ shown. Vertical lines show the respective AIC differences that were derived from the observed data.}
\end{figure}
\begin{figure}
\includegraphics[width=84mm, height=192mm]{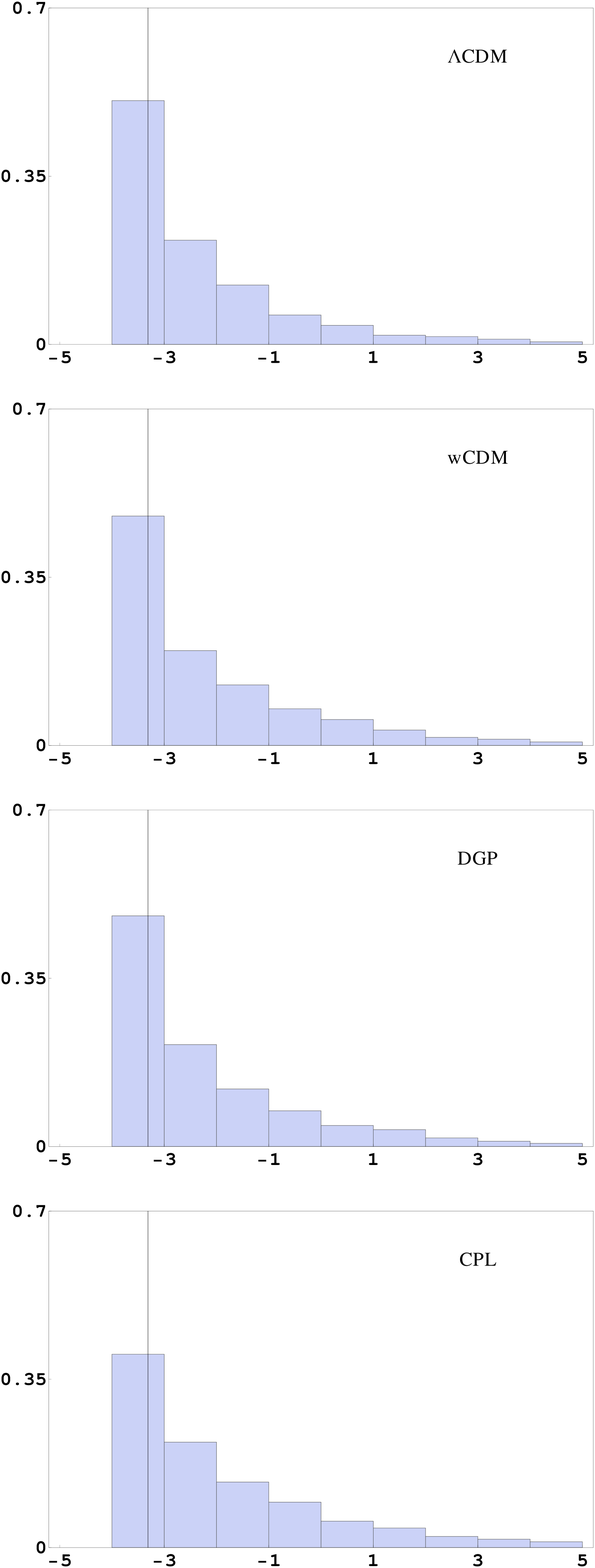}
\caption{~Probability distributions of the AIC differences between $\Lambda$CDM and CPL ($\Delta^d_{\Lambda CDM,CPL}$) for different reference model $C$. $C$ is used to generate the respective bootstrap samples and is written in the right upper corner of the figures. The horizontal axis indicates the $\Delta^d_{\Lambda CDM,CPL}$ value while the vertical axis indicates their relative frequency. If the process underlying our observations was really the best fit model of class $C$, then the values of the AIC differences under different realisations of noise would have the histogram distribution $\{\Delta^d_{\Lambda CDM,CPL} \ \vert \ d\in C_{371}\}$ shown. The exponential-like distributions observed are due to the fact that $\Lambda$CDM and CPL are nested models. Vertical lines show the respective AIC differences that were derived from the observed data.}
\label{CPLLCDMcomparison} 
\end{figure}
In order to intuitively understand what leads to these examples, it is 
instructive to consider the shapes of the distribution of AIC differences 
when comparing the $\Lambda$CDM and CPL models or the $\Lambda$CDM and DGP models. We note that a comparison of nested models will always involve 
strongly asymmetric, exponential-like distribution of AIC differences, 
while the comparison of non-nested models could result in almost symmetric
distributions. 
This is because the $\chi^2_{ML}$ for the general case cannot be smaller than the more specific case. For example, $\Lambda$CDM is a special case of the more general wCDM and CPL models. Hence, when we fit these models against 
any data or simulated data (regardless of the reference model used to 
generate it), the wCDM and CPL $\chi^2_{ML}$ values cannot be 
larger than the $\Lambda$CDM $\chi^2_{ML}$ value and are usually smaller. 
This results in a sharp edge along a maximum value of $\Delta_{\Lambda CDM, CPL}$  and an exponential-like distribution of AIC differences between 
these models.
\\
\\
We show examples of such a Gaussian-like distribution in 
Fig.~\ref{DGPLCDMcomparison1} and exponential-like distribution in 
Fig.~\ref{CPLLCDMcomparison}.  
As an aside, we note that the different 
reference models considered do not make much difference in the shape of 
these distributions in the figures. 
Since $\mbox{AIC} = \chi^2_{ML} + 2k$, the edge of these exponential-like histograms is shifted by twice the difference between their number of parameters. Hence for the
 comparison between the CPL model and $\Lambda$CDM model in Fig.~\ref{FailureRatevsthreshold} and Fig.~\ref{FailureRatevsthreshold-diff}, where the difference between the $\chi^2_{ML}$ is almost zero, the distribution of AIC differences $\Delta_{\Lambda CDM,CPL}$ has a sharp edge at approximately $4$. This implies that any AIC difference greater than a $\dt$ value of 
approximately $4$ will exclusively select the CPL model, irrespective of 
whether the reference model used was a CPL model or a $\Lambda$CDM model.
At lower values of $\dt$, the AIC selection procedure tends to select the model 
with the lower numbers of parameters since the $\chi^2_{ML}$ values are 
approximately the same. 
On the other hand, there is no similar constraint relating the $\chi^2_{ML}$ values
of the $\Lambda$CDM and DGP (at the best fit value); consequently the histogram turns out to be Gaussian-like. They have the same number of free parameters so this distribution is roughly centred about zero (the difference in $\chi^2_{ML}$ between these models). Obviously, if the quality of data was much better in terms of 
the error bars on each observation, or having a larger number of observations,
the differences in the $\chi^2_{ML}$ terms would be much larger
for the same choice of reference models used. In such cases, the model
comparison would be purely data-driven, deriving its discriminatory 
power from the `fit' term of AIC. This agrees with our intuitive idea that
a better dataset should be able to to resolve models better.

The shapes of the distribution of AIC differences are also important when we study the spread of the statistical uncertainty of the distribution of the AIC differences since the statistical spread of the distribution cannot be specified unless we know the shapes beforehand. Due to the structural differences between the two kinds of distributions, we must define the `error' bars according to the shape in order to make any useful comments about the uncertainty of the estimate. For the case of the Gaussian-like distribution, we define the error bars to be the standard deviation of the statistical distribution of differences. For the case of the exponential-like distribution, we define an error bar region as the range that begins at the sharp edge of the distribution and stops at the point where the range contains $68.3\%$ of the differences. We chose $68.3\%$, because 0.683 is approximately the probability of finding an outcome within a standard deviation of the mean of a Gaussian distribution.
\\
\\
As an illustration, we look at statistical uncertainty of the distribution of $\Delta^d_{\Lambda CDM,DGP}$ for the two separate cases where the DGP and $\Lambda$CDM models are the references. 
The distribution of the differences in both cases are Gaussian-like. 
We notice that the standard deviation of $\Delta^d_{\Lambda CDM,DGP}$ is 0.89 with DGP model as the reference
and 0.83 with $\Lambda$CDM model as the reference. The AIC difference between the DGP model and 
$\Lambda$CDM model $\Delta_{\Lambda CDM,DGP}$ in our original AIC analysis is 0.22 and smaller than the error bars. 
This result means that any subsequent analysis based on the value of $\Delta_{\Lambda CDM,DGP}$ observed is unreliable. 
The error bars could also be significant even if $\Delta_{\Lambda CDM,DGP}$ was larger than the error bars as we would have to 
modify any subsequent AIC difference analysis to include this uncertainty. 
When we look at $\Delta^d_{\Lambda CDM,CPL}$ (exponential-like distribution), the error bar region ranges from $-4.00$ to $-2.23$ for the case when the reference is the $\Lambda$CDM model and ranges from $-4.00$ to $-1.42$ when the reference is the CPL model. 
The $\Delta_{\Lambda CDM,CPL}$ value calculated from the empirical data was found to be $-3.31$. However, the error bar region range of approximately 2 would make the value of $-3.31$ less certain. Instead of quantifying the odds ratio given by Eqn.~\ref{odds} as having a value of 0.19, we now make a statement about its uncertainty by saying that the odds ratio can be a value between 0.14 and 0.48. These are just two examples in which one can carry out an analysis to determine the reliability of the model likelihood ratio $P(A)/P(B)$ that is calculated in Eqn. \ref{odds}. 
\\
\\
We note that the statistical uncertainty of the AIC differences obtained above is smaller than the $\dt$ value of 5 already mentioned above \citep{2007MNRAS.377L..74L,2007ApJ...666..716D,2009ApJ...703.1374S}. However, it is still significant enough and needs to be considered in our analysis.
\\
\\
\section{Conclusion}
\label{conclusion} 
AIC has been widely used as a technique for model selection. Most commonly,
 this has been applied by computing the AIC values for each candidate 
model through Eqn.~\ref{eqn:AIC} and selecting
the model with the smallest AIC value as the best model. 
The issue of considering the magnitudes of AIC differences between the 
models to indicate the relative plausibility or confidence in the models 
has also been addressed by Akaike and elaborated by 
Burnham and Anderson through the use of Akaike weights 
Eqn.~\ref{modellikelihood}. In the field of cosmology, AIC has been used in selecting 
models underlying the late time acceleration of the universe using data. 
There have also been suggestions of a rule of thumb that states an AIC difference 
of 5 or more would give strong evidence for the model with the smallest AIC value. This approximately corresponds to a ratio of model 
likelihoods of 12 or more. 
\\
\\
In this paper, we propose a method for calculating the ratio of the model likelihood between models $A$ and $B$ 
based on their AIC differences $\Delta_{A,B}$ (Appendix~\ref{app:Confusion}), using the idea of probabilistic model distinguishability by \citet{citeulike:7028761}. 
This is an alternative method of arriving at the odds ratio of Eqn.~\ref{odds} 
assuming that the AIC difference between the candidate models is a 
perfect unbiased estimator of the difference between the models' KL divergences with respect to the truth. 
A related analysis using the AIC differences in accordance with the Akaike weights 
also derives the same equation. The analysis of $\Delta_{A,B}$ was extended 
further by investigating the statistical uncertainty of this estimate. 
Our focus was not necessarily on the `best' cosmological theory. Thus, we did not use the most exhaustive data sets, 
nor explore in detail the systematics associated with the surveys considered.
\\
\\
To this end, we studied the distribution of the 
differences of AIC estimates given a certain quality of data $P(\Delta_{A,B} \vert E_N)$. 
Since we do not know the exact process underlying the empirical data ($E_N$ in Section \ref{sec:reliability}), 
we approach this problem by studying surrogate processes, where the 
reference or generating model is assumed to be one of four (best fit) candidate models for the late time acceleration of the universe; following the approach used by \citet{2007ApJ...666..716D}. 
These models were listed in Table.~\ref{tab:candidatemodels} with the best fit free parameters equal 
to the maximum likelihood values of each of these models. This was obtained 
by fitting 371 SNIa extracted from the Constitution compilation.
\\
\\
Our simulations have demonstrated that, given the data used, there was insufficient data to reliably use AIC to tell all the models apart; in agreement with the general consensus \citep{2007ApJ...666..716D}. For the case of $\dt=0$, the failure rate of the technique was shown to be particularly unsatisfactory. We also studied the reliability of the AIC technique when $\dt$ is increased to 2 and 5. Increasing $\dt$ results in increasing the number of cases where we cannot make a conclusion based on the AIC procedure. This was demonstrated by $\find$, 
which calculates the fraction of cases when the difference in AIC values between two models is less than $\dt$. We also studied $\ffalse,$ the proportion of cases where the AIC procedure using a threshold $\dt$ gives an incorrect result as a fraction of cases where we can make a conclusion (i.e. $\vert \Delta_{A,B}\vert > \dt$). We showed that $\ffalse$ does not necessarily 
decrease in the same universal way with an increasing $\dt$. Therefore, even when AIC chooses a model class (with a high level of 
$\dt$), the result is unreliable for at least some models within that 
model class. The demonstrated examples would perhaps not arise if the data 
was good enough that the differences in $\chi^2_{ML}$ was large. 
\\
\\
We also calculated the respective statistical uncertainty ($\sim 1 \ \sigma$ error bars) of $\Delta^d_{A,B}$ 
and showed it was even larger than the observed $\Delta_{A,B}$ between some of the models. This gives us a way to gauge the adequacy in the number of data points since the statistical uncertainty would become smaller than the observed differences when there is a sufficient amount of data. As an important example, we 
considered the $\Lambda$CDM and DGP models since they were shown to have the two lowest AIC values in Table.~\ref{tab:candidatemodels}. It was shown that the statistical uncertainty of $\Delta^d_{\Lambda CDM,DGP}$ was larger than the observed $\Delta_{\Lambda CDM,DGP}$, making it difficult to determine the better model between the two. From our simulation, we also showed that the shapes of the distribution of the AIC differences can be quite varied, 
ranging from a symmetric Gaussian-like distribution to an exponential-like distribution with a sharp edge and one sided tail. Thus, in order to use AIC reliably, one must pay proper attention to the statistical variation of $\Delta_{A,B}$.
\\
\\
In this paper, we made a number of assumptions to study the AIC technique. All calculations in this paper were only for an assumed reference. Since the empirical data does 
not give us $E_N$, there is no way to know the actual distribution of $\Delta_{A,B}$. However, we should note that AIC is a model comparison technique that assumes that one of the model 
classes contains the reference $C$. By restricting $C$ to the set of candidate models, we can at least look for statistical self-consistency in that assumption. It should be emphasised again that the reference models used were the best fit models and did not take into account the statistical uncertainty of the individual model parameters. That can be taken into account by sampling the distribution of parameters. As mentioned before, the whole point of this simulation is to highlight the statistical distribution of the AIC differences under different reference models and look for statistical inconsistencies under each of the assumptions. 
Another issue that should be noted is that the exact variance in the data is unknown and that the error bars in the data may not be reflective of the true error bounds. While we use these in our simulations, we note that the correctness of these error estimates was an assumption of 
previous AIC computations.
\\
\\
In summary, the reliability of the AIC 
estimator is an important issue that should be taken into consideration when 
using the AIC technique to select models. It should also be noted that such considerations 
are not just restricted to AIC but any technique that relies on the maximum likelihood estimators. 
This should be borne in mind when applying the techniques to any statistical analysis.

\section{acknowledgements}  
We would like to thank Yoshi Oono for checking the manuscript and providing many insights about the nature of AIC and the bootstrap technique. 
M.Y.J. Tan was funded in part by Yoshi Oono. During the course of this 
work, R. Biswas was partially supported by NSF AST 07-08849 and 
NSF AST 09-08693 ARRA and then by Argonne National Laboratory, operated by UChicago Argonne, 
LLC. Work at Argonne was supported by U.S. Department of Energy, Office 
of Science under contract DE-AC02-06CH11357. Finally, we would like to express gratitude to the anonymous 
reviewer whose input substantially improved the tone and direction of this paper.

\appendix
\section{Confusion Probability and Model Likelihood}
\label{app:Confusion} 
This is an almost identical repeat of \citet{citeulike:7028761} explanation of error probabilities which is framed in the language of hypothesis testing. It is reproduced for the convenience of the reader. Suppose $\{x_1,x_2,...,x_N\}\in \mathbb{X}^N$ are drawn independent and identically distributed (i.i.d) variables from one of $f_1$ and $f_2$ with $D(f_1\|f_2)<\infty$. Let $A_N \subseteq \mathbb{X}^N$ be the acceptance region for the hypothesis that the distribution is $f_1$ and define the type I and type II error probabilities as $\alpha_N = f_1^N(A_N^C)$ and $\beta_N = f_2^N(A_N)$ respectively. $A_N^C$ is the complement of $A_N$ in $\mathbb{X}^N$, and $f^N$ denotes the product distribution on $\mathbb{X}^N$ describing $N$ i.i.d outcomes drawn
from $f$. In this definition $\alpha_N$ is the probability that $f_1$ is mistaken for $f_2$, and $\beta_N$ is the probability of the opposite error. Stein's lemma tells us how low we can make $\beta_N$ given a particular value of $\alpha_N$. Indeed, let us define $\beta_N^{\epsilon} = \min_{A_N \subseteq \mathbb{X}^N , \alpha_N \leq \epsilon} \beta_N$ for a positive $\epsilon$. Then Stein's lemma tells us 
\begin{equation} 
\lim_{\epsilon \rightarrow 0} \lim_{N \rightarrow \infty}\frac{1}{N} \ln \beta_{N}^{\epsilon} = - D(f_1\|f_2). 
\end{equation} 

To prove Stein's lemma, we refer to the proof by \citet{citeulike:165404}, which is provided almost verbatim here for convenience sake. Defining $\delta \in \mathbb{R^{+}}$, we first state $A_N$ more explicitly as: 
\\ 
$A_N =$
\\
$\left\{ x \in \mathbb{X}^N : \exp[N \left(D(f_1\|f_2)-\delta \right)] \leq \frac{f_1(x)}{f_2(x)} \leq \exp[N \left(D(f_1\|f_2)+\delta \right)] \right\} $
Then, we have the following properties: 
\\ 
1. $f_1^N(A_N) \rightarrow$ 1. 
\\ 
Proof: 
\\
$f_1^N(A_N) = f_1^N \left(\frac{1}{N}\sum_{i=1}^N \log\frac{f_1(x_i)}{f_2(x_i)}\in  \left( D(f_1\|f_2)-\delta,D(f_1\|f_2)+\delta \right) \right)$
\\
$ \rightarrow 1 $ by the law of large numbers, since $D(f_1\|f_2)=E_{f_1}\left( \log \frac{f_1(x)}{f_2(x)} \right)$. Therefore, for any positive $\epsilon$, $\alpha_N < \epsilon$ for large $N$. 
\\ 
2. $f_2^N(A_N)\leq \exp\left[ -N\left(D(f_1 \| f_2) - \delta \right)\right]$\\ 
Proof: 
\begin{eqnarray*} 
f_2^N(A_N)&=& \sum_{A_N} f_2(x),\\ 
&\leq& \sum_{A_N} f_1(x) \exp[-N \left(D(f_1\|f_2)-\delta \right)],\\ 
&=& \exp[-N \left(D(f_1\|f_2)-\delta \right)] \sum_{A_N} f_1(x) , \\ 
&=& \exp[-N \left(D(f_1\|f_2)-\delta \right)] (1-\alpha_N). 
\end{eqnarray*} 
\\
\\ 
3. $f_2^N(A_N)\geq \exp\left[ -N\left(D(f_1 \| f_2) + \delta \right)\right]$ 
\\ 
Proof: 
\\ 
\begin{eqnarray*} 
f_2^N(A_N)&=& \sum_{A_N} f_2(x),\\ 
&\geq& \sum_{A_N} f_1(x) \exp[-N \left(D(f_1\|f_2)+\delta \right)],\\ 
&=& \exp[-N \left(D(f_1\|f_2)+\delta \right)] \sum_{A_N} f_1(x) , \\ 
&=& \exp[-N \left(D(f_1\|f_2)+\delta \right)] (1-\alpha_N). 
\end{eqnarray*} 
\\
\\ 
4. $\lim_{N\rightarrow \infty}\frac{1}{N} \log \beta_N = - D(f_1 \| f_2)$. 
\\ 
Proof: 
\\ 
From 2. and 3. we know: 
\begin{eqnarray*} 
\frac{1}{N} \log \beta_N \leq -D(f_1\|f_2)+\delta+\frac{\log(1-\alpha_N)}{N}.\\ 
\frac{1}{N} \log \beta_N \geq -D(f_1\|f_2)-\delta+\frac{\log(1-\alpha_N)}{N}. 
\end{eqnarray*}
\\
\\
5. No other sequence of acceptance regions does better. 
\\ 
Proof: 
Let $B_N \subseteq \mathbb{X}^N$ be any other sequence region with $\alpha_{N,B_N}=f_1^N(B_N^c)<\epsilon$. Let $\beta_{N,B_N}=f_2^N(B_N)$ 
\begin{eqnarray*} 
\beta_{N,B_N}&=& f_2^N(B_N),\\ 
&\geq& f_2^N(A_N \cap B_N),\\ 
&=& \sum_{A_N \cap B_N} f_2(x), \\ 
&\geq& \sum_{A_N \cap B_N} f_1(x) \exp[-N \left(D(f_1\|f_2)+\delta \right)], \\ 
&=& \exp[-N \left(D(f_1\|f_2)+\delta \right)] \sum_{A_N \cap B_N} f_1(x), \\ 
&\geq& (1-\alpha_N-\alpha_{N,B_N})\exp[-N \left(D(f_1\|f_2)+\delta \right)], 
\end{eqnarray*} 
where the last inequality is due to the following: 
\begin{eqnarray*} 
\sum_{A_N \cup B_N} f_1(x)&=& f_1(A_N \cap B_N),\\ 
&=& 1-f_1(A_N^c \cup B_N^c),\\ 
&\geq& 1-f_1(A_N^c)-f_1(B_N^c), \\ 
&=& 1-\alpha_N-\alpha_{N,B_N}. 
\end{eqnarray*} 
Hence, $\frac{1}{N} \log \beta_{N,B_N} \geq - D(f_1\|f_2)-\delta-\frac{\log(1-\alpha_N-\alpha_{N,B_N})}{N}$, and since $\delta >0 $, $\lim_{n\rightarrow \infty} \frac{1}{N} \log \beta_{N,B_N} \geq - D(f_1\|f_2)$. Thus no sequence of sets $B_N$ has an exponent better than $D(f_1\|f_2)$. 
\\
\\
In summary, property 1. shows that $A_N$ is the sequence that is generated by $f_1$ in the asymptotic limit. Properties 2., 3. and 4. derive the error probability of Stein's lemma and property 5. shows that $A_N$ is asymptotically optimal and the best error exponent is $D(f_1\|f_2)$.
\\
\\
Thus, we can interpret $\exp\left[-D\left(\mbox{truth} \| \mbox{model}\right)\right]$ as the probability of confusing the model with the truth or model probability, using the work of \citet{citeulike:7028761}.
\\
\\
This relation allows us to propose a slightly different but related way of interpreting the AIC difference between models as defining the ratio between model probabilities $\frac{P(A)}{P(B)}$ without resorting to Akaike weights. Let us start with 2 models $A$ and $B$ with $f_{A}$ and $f_{B}$ as their respective probability distribution functions. Their AIC values are $a$ and $b$ respectively and there is a difference of $\Delta_{A,B}=a-b$ between their AIC values. 

\begin{eqnarray*} 
\exp[-\Delta_{A,B}/2] &=& \frac{\exp[-a/2]}{\exp[-b/2]}, \\ 
&\approx& \frac{\exp\left[E_{X|\theta_0}[\log f_{A}(X|\hat{\theta}_A)]\right]}{\exp\left[E_{X|\theta_0}[\log f_{B}(X|\hat{\theta}_B)]\right]}, \\ 
&=& \frac{\exp\left[-D\left(\mbox{truth} \| f_{A}(X|\hat{\theta}_A)\right)\right]}{\exp\left[-D\left(\mbox{truth} \| f_{B}(X|\hat{\theta}_B)\right)\right]}, \\
&=& \frac{P(A)}{P(B)},
\end{eqnarray*} 
where $X$ represents data sampled from the truth $\theta_0$. $\hat{\theta}_A$ and $\hat{\theta}_B$ are the maximum likelihood parameters of model $A$ and $B$. 

\section{Bootstrap Method} 
\label{app:Bootstrap}
The bootstrap method used in this paper was a parametric bootstrap, where we made certain assumptions about the parametric relationship between the data input (explanatory variable) and the data output (response variable). We start with a set of $N$ actual data points $D=\{(x_1,y_1,\s_1),(x_2,y_2,\s_2),...,(x_N,y_N,\s_N) \}, $ where $y_i$ is the response variable that is observed with the error bar $\s_i$ and $x_i$ the explanatory (input) variable. To produce a bootstrap sample set $C_N$ consisting of $N$ data points, as noted in the text, we use a particular model $C$: $y = f(x)$ with the needed parameters chosen by maximum likelihood estimation. In the usual bootstrap approach, the obtained set $\{ y_i - f(x_i)\}$ is regarded as the estimate of the noise distribution, but in our case, unfortunately, the noise magnitude seems to depend on $x$. Therefore, we make an example of size $N$ bootstrap data as $\{f(x_i ) + \e_i \}$, where $x_i$ are the same as in $D$ and the noise $\e_i$ is a Gaussian random variable obeying $N(0,\s_i^2)$. With newly generated $\{\e_i\}$ we make a set of $N$ sample bootstrap data set $C_N$. For each bootstrap sample $d \in C_N$, we estimate the maximum likelihood parameters for model $A$ and $B$, respectively, and we can compute the set of AIC differences $\{ \Delta^d_{A,B} \ \vert \ d\in C_N\}$.  
 
\section{Getting $\chi_{ML}^2$ from SNIa, marginalising over $H_0$} 
\label{app:hubble}
We present the use of AIC in the context of cosmological model selection using SNIa from the Constitution compilation \citep{2009ApJ...700.1097H}. We fit the theoretical quantity of the distance moduli $\mu(z_{i})$ ($z$ is the observed red shift) against its observed value $\mu^{obs}_{i}=m_{i}-M_{i}$, where $m_{i}$ is the observed apparent magnitude and $M_{i}$ is the absolute magnitude of the Supernova data. Note that the index $i$ indicates the $i$th data point. 
\\
\\
$\mu(z)$ is calculated by the equation $\mu(z)= 5\log_{10} \frac{d_{L}(z)}{10pc}$, where $d_{L}(z)$ is the luminosity distance. We will assume that the universe is flat, by setting the curvature term $\Omega_{k}$ in the Hubble function $H(z)$ to zero, and under this assumption 
\begin{equation} 
d_{L}(z) = (1+z)c\int_{0}^{z}\frac{d z'}{H(z')}, 
\end{equation} 
where $c$ is the speed of light. 
\\
\\
We start with the assumption that $\mu^{obs}_{i}$ has a Gaussian noise structure. We model it as 
\begin{equation} 
\frac{1}{\prod_{i} \sqrt{2 \pi \s_{i}^{2}}} \exp \left[-\sum_{i}\frac{(\mu^{obs}_i-\mu(z_i))^2}{2 \s_{i}^2}\right], 
\end{equation} 
where $\s_{i}$ are consistent with the error bars associated with $\mu^{obs}_{i}$ in the Constitution compilation. 
Since $\mu^{obs}$ was calculated from the apparent (observed) magnitude $m$ by assuming a fixed absolute magnitude $M$ value which we are actually unsure about. We get around this problem by introducing a nuisance parameter $g$ and integrate it over a flat prior (Gaussian prior where the standard deviation $\rightarrow \infty$). To do this, we first integrate this over a Gaussian distribution of the nuisance parameter with standard deviation $\s_{g}^2$ to get 
\begin{eqnarray*} 
\int_{-\infty}^{\infty} \frac{1}{\prod_{i} \sqrt{2 \pi \s_{i}^{2}}} 
\exp \left[-\sum_{i}\frac{(\mu^{obs}_{i}-\mu(z_i)-g)^2}{2 \s_{i}^2}\right]
\\ 
\times \frac{1}{\sqrt{2 \pi \s_{g}^{2}}} \exp \left[-\frac{g^2}{2 \s_{g}^2}\right] dg. 
\end{eqnarray*}
\\
We can rewrite this in matrix form: 
\begin{eqnarray*}
\int_{-\infty}^{\infty} \frac{1}{\prod_{i} \sqrt{2 \pi \s_{i}^{2}}} \exp \left[-\frac{(X-gY)^{T}\Lambda(X-gY)}{2}\right] 
\\ 
\times
\frac{1}{\sqrt{2 \pi \s_{g}^{2}}} \exp \left[-\frac{g^2}{2 \s_{g}^2}\right] dg, 
\label{integral}
\end{eqnarray*}
where $X$ is a n-vector whose $i$th component is $\mu^{obs}_i-\mu(z_i)$, $Y$ is a n-vector where all the elements are `1s' and $\Lambda$ is the inverse of the covariance matrix (which in this case is diagonal). The $^T$ symbol denotes the transpose of a vector. Performing the $g$ integral and setting $\s_{g}$ to a large value, the following marginalised function is obtained: 
\begin{eqnarray*} 
\frac{1}{\sqrt{\s_g^2Y^{T} \Lambda Y}}\frac{1}{\prod_{i} \sqrt{2 \pi \s_{i}^{2}}} 
\exp \left[-\frac{1}{2}X^{T}\left( \Lambda - \frac{\Lambda Y Y^{T} \Lambda}{Y^{T} \Lambda Y}\right)X\right]. 
\end{eqnarray*} 
\\
This reduces the log likelihood to $\frac{X^{T} C X}{2}-\frac{1}{2}\log(\s_g^2Y^{T} \Lambda Y)-\sum_{i}\frac{1}{2}\log(2\pi \s_i^2)$. The second term suffers from a log divergence as $\s_g \rightarrow \infty$. However, since AIC works by comparing relative log likelihood values, we can regularise this term away by setting it to zero. We can also ignore the third term since it is a fixed constant independent of the parameter choice. 
\\
\\
This reduces finding the maximum likelihood to minimising $X^{T} C X$, where $C=\Lambda - \frac{\Lambda Y Y^{T} \Lambda}{Y^{T} \Lambda Y}$. Because of the marginalisation against a flat prior, the rank of $C$ is smaller than the rank of $\Lambda$ by one and thus $C$ cannot be inverted. The marginalisation procedure also implies that the choice of the Hubble parameter $H_0$ and even the speed of light $c$ is irrelevant to finding the maximum likelihood values of the other parameters.
\\
\\
This leads to the following relative AIC term: 
\begin{equation} 
\mbox{AIC}=X^{T}(\hat{\theta}) C X (\hat{\theta}) + 2 k , 
\end{equation} 
where $\hat{\theta}$ is the set of parameters that minimises $X^{T} C X$. 
\\
\\
The first term corresponds to the maximum likelihood while the second term is the bias correction which is dependent on the number of free parameters. The maximum likelihood parameters can then be found using some common minimisation procedure. Specifically, these were found using the Gauss-Newton algorithm \citep{bjor:96}. This allows us to calculate the AIC values for 4 candidate models. The data used consists of 371 Supernova events taken from the Constitution compilation (MLCS table) \citep{2009ApJ...700.1097H}. We computed the AIC values for the different models and found that the DGP model has the smallest AIC value among the four models we considered in this paper. 
\\
\\
As a very small technical side issue, it should be noted that unlike the other work \citep{2005PhLB..623...10G} that used AIC as a model selection tool, as described in the above, we marginalised away the $H_0$ term against a flat prior which reduces the number of free parameters by one. This technical difference alone should not affect our use of AIC.

\bibliographystyle{mn2e} 
\bibliography{AIC_cosmology}
\end{document}